\documentclass[aps,prl,twocolumn,groupedaddress,showpacs]{revtex4}
\usepackage{graphicx}

\begin{document}
\title{Quantum reflection from a solid surface at normal incidence}
\author{T.~A. Pasquini, Y. Shin, C. Sanner, M. Saba, A. Schirotzek, D.~E. Pritchard, W. Ketterle}
\homepage{http://cua.mit.edu/ketterle_group/}

\affiliation{Department of Physics, MIT-Harvard Center for
Ultracold Atoms, and Research Laboratory of Electronics,
Massachusetts Institute of Technology, Cambridge, Massachusetts,
02139}
\date{\today}

\begin{abstract}

We observed quantum reflection of ultracold atoms from the
attractive potential of a solid surface. Extremely dilute
Bose-Einstein condensates of $^{23}$Na, with peak density
$10^{11}-10^{12}$~atoms/cm$^3$, confined in a weak
gravito-magnetic trap were normally incident on a silicon
surface.  Reflection probabilities of up to $20\%$ were observed
for incident velocities of $1-8$~mm/s. The velocity dependence
agrees qualitatively with the prediction for quantum reflection
from the attractive Casimir-Polder potential. Atoms confined in a
harmonic trap divided in half by a solid surface exhibited
extended lifetime due to quantum reflection from the surface,
implying a reflection probability above 50\%.

\end{abstract}


\maketitle

Quantum reflection is a process in which a particle reflects from
a potential without reaching a classical turning point. A solid
surface provides a long-range attractive potential for atoms. At
separation, $r$, much larger than the atomic radius the potential
follows the Casimir-Polder expression
$U=-C_4/((r+3\lambda/2\pi^2)r^3) $, where $\lambda$ is the
effective atomic transition wavelength~\cite{CAS48}. Classically,
an atom incident on such a potential will be accelerated toward
the surface, resulting in inelastic scattering or adsorption. A
quantum mechanical treatment of an atom-surface collision reveals
that the atom is reflected from the purely attractive surface
potential if the potential energy changes abruptly on a length
scale set by the quantum mechanical wavelength~\cite{CLO92, CAR98,
MODY01, FRI02}. The condition for significant reflection is that
the local particle wavenumber normal to the surface,
$k_\bot=\sqrt{k_{\infty}^2-2mU/\hbar^2}$, change by more than
$k_\bot$ over a distance $1/k_\bot$ . Here,
$k_\infty=mv_\bot/\hbar$ is the normal wavenumber of the atom far
from the surface, $m$ is the atomic mass, $v_\bot$ is the normal
incident velocity, and $\hbar$ is the Planck constant divided by
$2\pi$. The reflection probability, $R$, tends to unity as the
incident velocity tends to zero, $R\approx
1-4\beta_4mv_\bot/\hbar$, where $\beta_4$ is the length scale
associated with the $C_4$ coefficient, $C_4=\beta_4^2\hbar^2/2m$.
High probability quantum reflection requires low incident velocity
or weak attraction to the surface, conditions previously realized
only in exceptional systems.

Studies of quantum reflection were first performed with helium or
hydrogen atoms incident on liquid helium surfaces~\cite{NAY83,
BER89, DOY91, YU93}.  The extremely weak interaction strengths and
low mass atoms allowed for quantum reflection at relatively high
incident energies of $\sim k_B \times 10$ mK~\cite{NAY83, YU93},
where $k_B$ is the Boltzmann constant. Reflection of noble and
alkali atoms from a solid surface requires that atoms be incident
with a million times less energy, $\sim k_B \times 10$ nK. This
has been accomplished only by letting untrapped atoms hit solid
surfaces at grazing incidence~\cite{AND86, SHI01, SHI02a, SHI02b,
DRU03}, meaning that most of the velocity is directed parallel to
the surface and reflection only deflects atoms slightly.
Reflection probabilities in excess of 60\% have been observed for
incidence angles of a few milliradians~\cite{SHI02a}. Atom-surface
potentials have also been studied in the presence of evanescent
light waves generated by total internal reflection at a glass
surface~\cite{LAN96, SEG97}.

Normal-incidence quantum reflection of trapped atoms may be
exploited in the construction of novel atom optical devices. The
current generation of atom mirrors for reflecting, confining and
focusing ultracold atoms employs evanescent waves produced by
total internal reflection of laser light~\cite{DOW96} or
magnetized materials~\cite{HIN99}. An atom-optical device based on
quantum reflection is in a category of its own, as it works using
the universal atom-surface interaction and depends on the long
wavelength of ultracold atoms. Past experiments with grazing
incidence atomic beams have demonstrated a mirror~\cite{SHI01}, a
reflective diffraction grating~\cite{SHI02a}, and a hologram based
on quantum reflection~\cite{SHI02b}.

In this Letter, we demonstrate normal-incidence quantum reflection
of ultracold sodium atoms. Using the harmonic trapping potential
of a gravito-magnetic trap~\cite{LEA03a, MON92}, we varied the
center of mass velocity of dilute Bose-Einstein condensates and
induced controlled collisions with a silicon surface at velocities
as low as 1 mm/s, corresponding to collision energies of $k_B
\times 1.5$ nanokelvin or $1.2 \times 10^{-13}$ eV.  A reflection
probability of $\sim 20\%$ was obtained for an incident velocity
of 2 mm/s, realizing an atom mirror. Our experimental results are
in qualitative agreement with theoretical predictions for single
atoms incident on a conducting surface. Additionally, atoms were
confined in one dimension by a silicon surface, where lifetime
measurements indicate reflection probabilities in excess of
$50\%$.

Bose-Einstein condensates of $^{23}$Na atoms were prepared and
transferred into a gravito-magnetic trap, comprising a single coil
and three external bias fields, as described in
Ref.~\cite{LEA03a}. Mounted 1 cm above the center of the single
coil was a $\sim$20 $\mu$m thick N-type doped polished Si (100)
surface with a resistivity of 1 - 10 $\Omega$-cm. For typical
loading parameters, condensates containing $3\times10^5$ atoms
were confined 200 $\mu$m to one side of the surface in a harmonic
trap characterized by $(\omega_\bot, \omega_y, \omega_z)=2\pi
\times (10, 10, 6.5)$ Hz, where $\omega_\bot$ is the horizontal
trap frequency perpendicular to the surface, $\omega_y$ is the
horizontal trap frequency parallel to the surface, and $\omega_z$
is the vertical trap frequency. At this point, $\omega_\bot$ and
$\omega_y$ were adjusted between 11 Hz to 2 Hz by changing the
vertical bias field as described in Ref.~\cite{LEA03a}.  The
position of the trap center relative to the surface was controlled
by applying a bias field, $B_\bot$, perpendicular to the
surface~\footnote{The trap center to surface separation exhibited
a linear dependance on $B_\bot$.  The relative position of the
surface was determined by loading a condensate into the trap,
slowly moving the trap center toward the surface, holding for
several trap periods, and measuring the remaining atom number. As
the separation became smaller than the condensate radius, atoms
were lost from the trap due to adsorption on the surface. The
surface position was the point at which no atoms remained in the
trap, and was in good agreement with the visual location of the
surface determined by imaging.  The position of the surface was
determined to within a 10 $\mu$m.}. Empirically, we find that near
the surface the non-condensed fraction of the atomic cloud is
reduced by the ``surface evaporation" effect, in which adsorption
preferentially removes the hottest atoms from the
cloud~\cite{LUI93,HAR03}.

\begin{figure}[t]
\begin{center}
\includegraphics{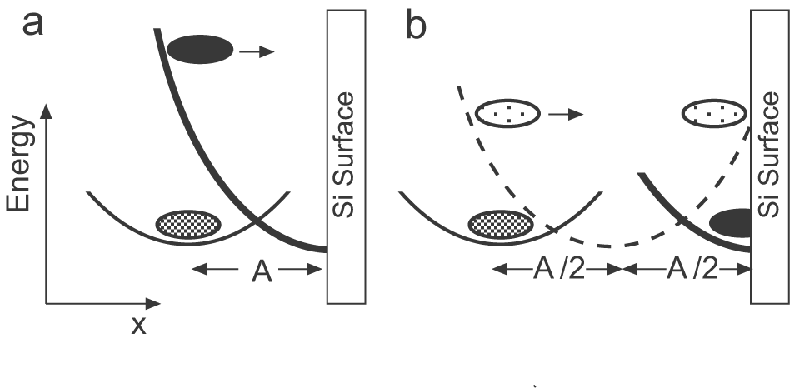}
\caption{Experimental schematic. Atoms were confined in a
gravito-magnetic trap~\cite{LEA03a} with trap frequencies ranging
from $2\pi \times (2, 2, 6.5)$ Hz to $2\pi \times (11, 11, 6.5)$
Hz, near a $\sim 20$ $\mu$m thick Si surface. (a) Quantum
reflection was studied by inducing a dipole oscillation of
amplitude A perpendicular to the surface and centered on the
surface. The incident velocity was varied from 1-8 mm/s by
adjusting $A$. (b) Atoms were loaded into a surface trap with zero
incident velocity using an intermediate trap located at $A/2$.
Atoms initially confined at a distance $A$ from the surface were
made to undergo a dipole oscillation of amplitude $A/2$ by
shifting the trap center halfway to the surface. After holding for
half a trap period, $T_\bot/2$ , the atoms were incident on the
surface with near zero velocity. The trap center was again shifted
by $A/2$ towards the surface, trapping the atoms against the Si
wafer. To ensure contact between atoms and the surface, the center
of the final trap was located $\sim10\%$ of the original
condensate size beyond the Si surface. \label{f:schematic}}
\end{center}
\end{figure}

\begin{figure}
\begin{center}
\includegraphics{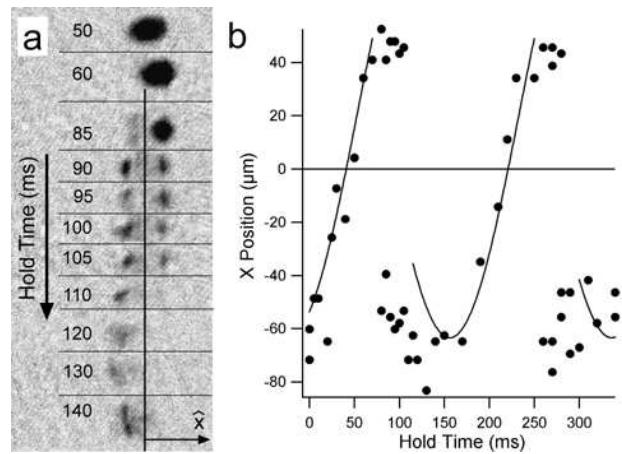}
\caption{Atoms reflecting from a Si surface.  Atoms confined
$\sim$70 $\mu$m from a Si surface were transferred into a harmonic
trap centered on the surface with $2\pi \times (3.3, 2.5, 6.5)$
Hz. The dipole oscillation of the condensate was interrupted
periodically by collisions with the surface, which reversed the
cloud's center of mass velocity. After a variable hold time, atoms
were released from the trap, fell below the edge of the surface
and were imaged with resonant light after 26 ms time-of-flight.
The position of the atoms in time-of-flight is the sum of the
in-trap position at the time of release and the product of the
release velocity and time-of-flight. (a) Time-of-flight images of
atoms after increasing hold times show the partial transfer of
atoms from the initial condensate (right) into the reflected cloud
(left) as the collision occurs. The separation is due to the
reversed velocity. The vertical line shows the horizontal location
of the surface.  The field of view is 1.4 mm wide. (b) The
time-of-flight positions of the incident and reflected atom clouds
relative to the surface are well modeled by a single particle
undergoing specular reflection in a half harmonic trap (solid
line). During collision, the behavior deviates from the single
particle model because of the finite cloud diameter of $\sim$60
$\mu$m.  A second reflection is visible at 270 ms.
\label{f:bounce}}
\end{center}
\end{figure}

The dipole mode of a harmonically trapped condensate is identical
to the behavior of a pendulum; atoms oscillate with amplitude $A$
and trap period $T_{\bot}=2\pi/\omega_\bot$. The presence of a
surface within the trapping potential dramatically alters the
dipole oscillation in the same way a wall would alter the
oscillation of a pendulum. Bose-Einstein condensates undergoing
dipole oscillation in the gravito-magnetic trap were made to
collide with the solid silicon surface as described in
Figure~\ref{f:schematic}a. Collision with the surface occurred at
time $\tau_C\approx T_\bot/4$ with incident velocity
$v_\bot=A\omega_\bot\approx 1.5$ mm/s. This phenomenon is observed
in Figure~\ref{f:bounce}. Near $\tau_C$, two distinct velocity
classes were visible corresponding to atoms in the initial
condensate and atoms reflected from the surface. The simultaneous
presence of incident and reflected atoms is explained by the fact
that the back of the condensate hits the surface $\sim$40 ms after
the front end due to the $\sim$60 $\mu$m condensate diameter and
slow (1.5 mm/s) velocity.  At collision, the harmonic motion of
the atoms was phase shifted by $2(\pi-\omega_\bot\tau_C)$, as seen
in Figure~\ref{f:bounce}b.

The reflected atom cloud was smaller than the incident one and had
a comparable density.  In some instances, the cloud appeared to
have a bimodal distribution, indicating that coherence might be
preserved in the collision.  The reflected atoms continued to
oscillate in the trap with the original amplitude, suggesting that
atoms reflected specularly and that the kinetic energy was
conserved during the collision. Eventually, the reflected atom
cloud underwent a second collision with the surface at $\sim
T_\bot/2$ after the first collision. Additional collisions were
not observed as the atom number fell below our detection limit.

\begin{figure}
\begin{center}
\includegraphics{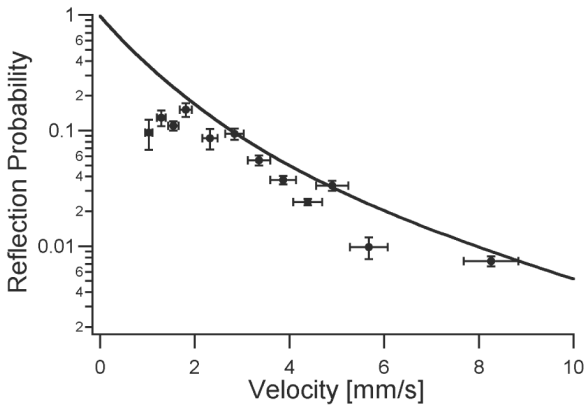}
\caption{Reflection probability vs incident velocity.  Data were
collected in a magnetic trap with trap frequencies $2\pi \times
(3.3, 2.5, 6.5)$ Hz.  Incident and reflected atom numbers were
averaged over several shots. Vertical error bars show the standard
deviation of the mean of six measurements. Horizontal error bars
reflect the uncertainty in deducing $v_\bot$ from the applied
magnetic field $B_\bot$. The solid curve is a numerical
calculation for individual atoms incident on a conducting surface
as described in the text. \label{f:reflective}}
\end{center}
\end{figure}

We observed reflected fractions that varied from 0 - 20$\%$ over
the incident velocity range of 1 - 8 mm/s, corresponding to a
collision energy of $\sim k_B \times (1-100)$ nK, for atoms with a
peak density of $\sim 2 \times 10^{12}$ cm$^{-3}$ in a $2\pi
\times(3.3, 2.5, 6.5)$ Hz trap. Figure~\ref{f:reflective} shows
the measured reflection probability, defined as the ratio of
reflected atom number to incident atom number, as a function of
incident velocity. The reflection probability increases with
decreasing velocity, a signature of quantum reflection. Similar
behavior was observed for atoms with a peak density of $\sim 7
\times 10^{12}$ cm$^{-3}$ in a $2\pi \times(10.5, 10, 6.5)$ Hz
trap.  For comparison, we include a line in
Figure~\ref{f:reflective} showing the calculated reflection
probability for a single atom incident on an conducting wall.  The
reflection probability was obtained by numerically solving the
Schr\"{o}dinger equation with the Casimir-Polder potential using
the $C_4$ coefficient calculated for sodium atoms incident on a
conducting surface, $C_4=9.1\times 10^{-56}$ Jm$^4$~\cite{CAS48}
and $\lambda=590$ nm. This model ignores the harmonic trapping
potential, inter-atomic forces, and electrostatic effects of
adsorbed alkali atoms on the surface, which have recently been
shown to distort the long range potential in the case of Rb atoms
on insulating surfaces~\cite{MCG04, LIN04}. Furthermore, the doped
Si surface has a finite conductivity, leading to a reduction in
$C_4$ of order $20-40\%$ and a slightly higher reflection
probability than a perfect conductor~\cite{YAN97}.

According to the model, reflection of atoms with 2 mm/s velocity
occurs at a distance of $\sim$1 $\mu$m from the surface, where the
full potential is approximated to within 10$\%$ by $U=-C_4/r^4$.
The range of velocities we could explore is not large enough to
investigate the region closer to the surface where the potential
has a $1/r^3$ dependence.  It should be noted that without
retardation, the reflection probability would be more than a
hundred times lower. Ultimately, quantum reflection may be a
powerful tool to characterize atom-surface interactions.

We also observe dynamics, not present in single-atom quantum
reflection, when a Bose-Einstein condensate is incident on a
surface.  For incident velocity below 2 mm/s, the measured
reflection probability remained approximately constant between 10
and 15\%, in contrast with theoretical predictions and previous
observation. This discrepancy may be due to collective excitations
of the atoms or acceleration from the harmonic trapping potential
during the collision. The shape and density of the reflected atom
cloud, as can be seen in Figure~\ref{f:bounce}a, were not
reproducible from shot to shot. Reflected clouds were excited and
sometimes fragmented and higher velocity incident atoms tended to
produce more dense reflected clouds, which may imply that an
excitation occurred during the collision that was more pronounced
at low collision velocities.

Furthermore, the role of the mean-field interaction energy should
be considered. When a condensate is released from a trapping
potential, the repulsive mean-field energy is converted into
kinetic energy, imparting to the atoms an average velocity equal
to the local speed of sound, $c=\sqrt{gn/m}$, where
$g=4\pi\hbar^2a/m$ is the coupling parameter associated with
atom-atom interaction, and $a$ is the s-wave scattering length. We
expect that the mean-field energy will be released during the
collision so that, even for a condensate with zero center of mass
velocity, the incident velocity may be characterized by the speed
of sound. For Na condensates at a density of $2\times 10^{12}$
cm$^{-3}$, this velocity is $\sim 0.6$ mm/s.

\begin{figure}
\begin{center}
\includegraphics{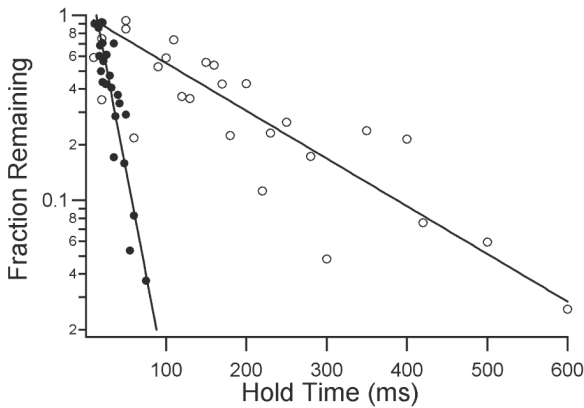}
\caption{Lifetime in the Si surface trap. Solid (open) circles
show the remaining atom fraction vs time for a $2\pi \times (9, 9,
6.5)$ ($2\pi \times (3.3, 2.5, 6.5)$) Hz trapping potential with
an initial atom number $3\times 10^4$ ($9\times 10^4$).  The solid
line exponential fit gives a lifetime of 23 ms (170 ms) for the
high (low) frequency trap geometry. The lifetime for atoms
confined far from the surface exceeded 20 s for either geometry.
\label{f:lifetime}}
\end{center}
\end{figure}

In the limit of zero incident velocity, a surface acting as ideal
atom mirror could be used to construct a physical container for
ultracold atoms and Bose-Einstein condensates. To examine the
feasibility of confining atoms with solid surfaces, atoms were
held in a magnetic trap divided in half by the Si surface. The
transfer procedure is described schematically in
Figure~\ref{f:schematic}b.  Figure~\ref{f:lifetime} shows the
remaining fraction of atoms in the trap as a function of hold time
for two different magnetic trap parameters, one at high trap
frequency, $2\pi \times(9, 9, 6.5)$, and the other at low trap
frequency, $2\pi \times(3.3, 2.5, 6.5)$. After an initial loss due
to the non-zero incident velocity (not shown), the atom number was
found to decrease exponentially.  The lifetime for the high (low)
frequency trap was 23 ms (170 ms). We attribute the losses of
atoms to scattering with the surface and adsorption. Fluctuating
electro-magnetic fields in a (semi)conductor can also induce
losses of atoms due to thermally induced spin flips (see, e.g.,
Ref.~\cite{LIN04}). However, at the large magnetic bias field of
$\sim$10 G, atoms can be ejected from the trap only with fields of
frequency $\sim$7 MHz. At an average distance of $\sim$15 $\mu$m
from surface, the spin flip decay rate should not exceed 0.1
$\mu$Hz, a negligible effect in the present experiment and not a
significant limitation for future ones.

In order to estimate the effect of quantum reflection on the
lifetime of atoms in the trap, we assume that the atoms are
incident on the surface with a velocity proportional to the speed
of sound in the condensate and that the geometry of the trapped
atom cloud is independent of the trap frequency. The atom loss
rate to the surface may be expressed as: $dN/dt\propto -nSc(1-R)$
where $S$ is the contact area between surface and condensate. From
this rate equation, we express the lifetime as
$\tau_L=-N/(dN/dt)=\chi T_\bot/(1-R)$, where $\chi$ is an
undetermined numerical parameter independent of the trap
frequency.  An identical equation would describe a thermal cloud
of atoms.  Comparing the ratio $\tau_L/T_\bot$ for the two
different trap frequencies, we cancel out the constant $\chi$.
Assuming the reflection probability for the high-frequency trap,
$R_h = 0$, gives a value of $R_l = 60\%$ for the low-frequency
trap reflection probability.  A more reasonable assumption of $R_h
= 20\%$ gives a value of $R_l = 70\%$.

The results presented here demonstrate that large quantum
reflection probability is not confined to exotic surfaces or
extreme angles of incidence: a simple silicon wafer at room
temperature can function as an atomic mirror at normal incidence,
reflecting ultracold atoms. The construction of practical atom
optical devices based on normal-incidence quantum reflection
requires even higher reflection probabilities than demonstrated in
this work. Such improvements are predicted for low-density and
extremely thin surfaces~\cite{MODY01}, and have been observed with
patterned surfaces, where a reduction in surface density by
etching increased the maximum reflection probability by a factor
of two~\cite{SHI01}. Because reflection occurs far from the
surface, uniformity of the surface is not a critical factor, as
roughness is averaged over the atomic wavelength.

Surfaces are traditionally considered enemies of cold atoms: laser
cooling and atom optics have developed thanks to magnetic and
optical traps that confine atoms with non-material walls in
ultra-high vacuum environments designed to prevent contact with
surfaces. Paradoxically, it turns out that in the extreme quantum
limit of nanokelvin matter waves, a surface at room temperature
might become a useful device to manipulate atoms.

This work was supported by NSF, ONR, ARO, DARPA, and NASA.  We
thank A.E. Leanhardt for insightful discussion and J.M. Doyle for
a critical reading of the manuscript.  M.S. acknowledges
additional support from the Swiss National Science Foundation.
C.S. acknowledges the support of the Studienstiftung des deutschen
Volkes.

\end{document}